Tech Science Press

# Deployment of Polar Codes for Mission-Critical Machine-Type Communication Over Wireless Networks

**Najib Ahmed Mohammed[1], Ali Mohammed Mansoor[1,*], Rodina Binti Ahmad[1] and Saaidal Razalli Bin Azzuhri[2]**

[1]Department of Software Engineering, Faculty of Computer Science and Information Technology, University of Malaya, 50603, Kuala Lumpur, Malaysia

[2]Department of Computer System & Technology, Faculty of Computer Science and Information Technology, University of Malaya, 50603, Kuala Lumpur, Malaysia

*Corresponding Author: Ali Mohammed Mansoor. Email: ali.mansoor@um.edu.my

Received: 25 May 2021; Accepted: 03 August 2021

**Abstract:** Mission critical Machine-type Communication (mcMTC), also referred to as Ultra-reliable Low Latency Communication (URLLC), has become a research hotspot. It is primarily characterized by communication that provides ultra-high reliability and very low latency to concurrently transmit short commands to a massive number of connected devices. While the reduction in physical (PHY) layer overhead and improvement in channel coding techniques are pivotal in reducing latency and improving reliability, the current wireless standards dedicated to support mcMTC rely heavily on adopting the bottom layers of general-purpose wireless standards and customizing only the upper layers. The mcMTC has a significant technical impact on the design of all layers of the communication protocol stack. In this paper, an innovative bottom-up approach has been proposed for mcMTC applications through PHY layer targeted at improving the transmission reliability by implementing ultra-reliable channel coding scheme in the PHY layer of IEEE 802.11a standard bearing in mind short packet transmission system. To achieve this aim, we analyzed and compared the channel coding performance of convolutional codes (CCs), low-density parity-check (LDPC) codes, and polar codes in wireless network on the condition of short data packet transmission. The Viterbi decoding algorithm (VA), logarithmic belief propagation (Log-BP) algorithm, and cyclic redundancy check (CRC) successive cancellation list (SCL) (CRC-SCL) decoding algorithm were adopted to CC, LDPC codes, and polar codes, respectively. Consequently, a new PHY layer for mcMTC has been proposed. The reliability of the proposed approach has been validated by simulation in terms of Bit error rate (BER) and packet error rate (PER) *vs.* signal-to-noise ratio (SNR). The simulation results demonstrate that the reliability of IEEE 802.11a standard has been significantly improved to be at PER = $10^{-5}$ or even better with the implementation of polar codes. The results also show that the general-purpose wireless networks are prominent in providing short packet mcMTC with the modification needed.

## 1 Introduction

Machine-type Communication (MTC) is an emerging vision which is featured by the exchange between intelligent devices, processing, actuation, and fully automatic data generation with or without low human intervention [1]. Unlike human-type communication (HTC) which demands large data amount at high data rates, most MTC devices transmit only tens of bytes of data and guarantee ultra-high reliability besides providing very low latency, which have been neglected in current general purpose wireless networks. MTC, also called Machine to Machine (M2M) communication, currently plays an imperative role in wireless communication and cellular networks. It is being applied in different platforms such as Vehicle-to-vehicle (V2V) communication, Device-to-device (D2D) communication, Device-to-everything (D2X) communication, Industrial Internet of Things (IIoT), and fourth generation industrial (Industrial 4.0) communication. Thanks to the rapid advancement of embedded device, MTC has become the controlling communication paradigm for real time applications such as factory automation, monitoring systems, remote manufacturing, smart metering, healthcare [2], utilities, transportation, and numerous other applications [3]. While some of these real-time applications are served through the infrastructure of wired networks, it is possible to open their actual perspectives just by adopting wireless networks. Wireless networks can bring numerous benefits to MTC such as flexibility, design simplification, and cost reduction compared to the currently used wired infrastructure [4]. Despite that, wireless communications pose significant threats to reliability performance which is prone to frequency packet loss due to moving objects, electromagnetic fields, and metal surfaces [5]. Therefore, recent wireless networks are incapable of addressing the stringent requirements of most mission critical MTC (mcMTC) applications. mcMTC applications require ultra-high reliability with packet error rate (PER) of $10^{-5}$ or even better, very low latency goes to 1 ms or even lesser, and ubiquitous communication for supporting short packet MTC, whereby the consequences of service failure are severe.

Numerous research challenges have been reported recently while adopting wireless networks to realize mcMTC requirements [6]. They are based on IEEE 802.15 standards, IEEE 802.11 standards, or cellular networks [7] and each method has pros and cons. However, none of these networks satisfies the stringent requirements of mcMTC due to the Physical (PHY) layer overheads [8] and inefficiencies in channel coding [9]. While IEEE 802.15 and IEEE 802.11 standards possess the features of low mobility, small data transmission, group-centric communications, and others to support short packet transmission and high data rates [10,11], their PHY layer limitations prevent them from providing mcMTC services. From the cellular networks perspective, the current LTE technology is incapable of supporting mcMTC services due to its resource blocks' structure, which requires 1 ms time transmission interval (TTI) with high orthogonal frequency division multiplexing (OFDM) symbols (14 symbols size). Even though 5G promises mcMTC, the network overheads incurred in the underlying cellular infrastructure affects latency which is expected to be no lower than 1 ms. Due to the aforementioned reasons, wired real-time Ethernet networks like EtherCAT currently provide MTC [12].

In general, the packet length in MTC is very short [13–15], and an efficient channel coding scheme is paramount to provide ultra-reliable MTC [16,17]. Channel coding is a forward error correction (FEC) method where the message is redundantly encoded using error correction codes by the sender such that the receiver can correct the errors of the data transmitted over distorted or noisy communication channels. This method reduces the need for retransmission of data, hence reducing channel delay. Due to this channel coding capability, it has become an essential part of the PHY layer of current wireless standards.

In this paper, we have proposed a new PHY layer design based on IEEE 802.11 standard. It aims at providing ultra-reliability to be at $PER = 10^{-5}$ or even better within latency bound by implementing an efficient channel coding scheme for short packet length in order of a few hundred bits. To achieve this, we first reviewed the state-of-the-art of mcMTC deployment over wireless networks, their challenges and proposed solutions. Afterward, we studied the possibility of IEEE 802.11 standard in providing short packet mcMTC. Subsequently, we investigated the reliability performance of modern channel coding schemes in short block regime. Building on this study, we outlined the primary directions for developing mcMTC systems over wireless networks.

The remaining parts of this research are structured as follows. Section 2 reviews the motivation and related works. Section 3 and Section 4 elaborate the potential latency and reliability solutions for short packet mcMTC respectively. The proposed PHY layer design has been provided in Section 5 and the evaluation has been given in Section 6. Finally, Section 7 provides the conclusion and future work.

## 2 Motivation and Related Work

This section reviews the cutting-edge mcMTC deployment through wireless networks, the challenges in adopting mcMTC over current wireless standards, and the channel coding techniques applied in the PHY layer considering short packet communication.

The most challenging features of mcMTC are the stringent requirements in providing ultra-high reliability, very low latency, and concurrent transmission for very large number of devices under the condition of short packet transmission. Therefore, adopting an existing wireless network to work as a function for MTC has been extensively studied to overcome these challenges. While the foundation of MTC has been provided in the work of Polyanskiy et al. [18], the growth of wireless sensors for industrial utilization like the IWSNs [19] pushes MTC into industrial sectors where they are being adopted in several mcMTC deployments [4]. Many standards have been developed which include Wireless Highway Addressable Remote Transducer (WirelessHART) [20], ISA100.11a [21], Wireless Networks for Industrial Automation for Process Automation (WIA-PA) [10], and Wireless Interface for Sensors and Actuators (WISA) [22]. They are deployed on top of IEEE 802.15.4 PHY layer. On the other hand, the IEEE 802.11 standard has been considered as a foundation for MTC in many standards such as the real time WiFi (RT-WiFi) standard [11], wireless networks for industrial automation-factory automation (WIA-FA) standard [23], and Wireless High Performance (WHP) [24] proposed solution for MTC which has been improved in [25,26]. However, most of these general-purpose wireless developments are based on customizing upper layers while keeping PHY and sometimes MAC layers untouched. Such standards can allow easier inter-compatibility and faster standardization. However, they present a network performance limitation towards mcMTC as all these standards are restricted to the performance of their PHY base work. i.e., IEEE 802.15.4 and IEEE 802.11 PHY layers.

From cellular networks perspective, in the current LTE standard, the channel estimation overhead takes about 5 ms, and the exclusive metadata reserved further lowers transmission rate. However, even though 5G promises MTC, many real-time applications, like closed-loop factory automation controls necessitating large refresh rates or periodic mcMTC, are out of the 5G scope [27]. There are also doubts that factory operators can depend on outside mobile telecommunication providers for their onsite applications [4]. Based on the reasons mentioned above, all the proposed solutions have their own limitations to satisfy the requirements demanded by most mcMTC scenarios.

The mcMTC has a significant technical impact on the design of all layers of the communication protocol stack. However, mcMTC requires innovative PHY layer design considering efficient modulation scheme, handling of inter-symbol interference (ISI), providing pilot transmission strategy [28,29], and efficient channel code [30]. Latency reduction has been extensively studied in [8] as it provided a very short preamble length made of only one OFDM symbol compared to five symbols occupied in the IEEE 802.11a/g standard. Though the study experimental results are optimistic towards minimizing the overall latency, reliability has been neglected as channel coding has not been considered in its experimentation. On the other hand, the channel coding design has a crucial impact on the reliability performance, especially for short blocks causing a reduction in channel observations, consequently increasing the gap to the Shannon limit, i.e., for short block codes, Shannon's theoretical model fails [15]. In general, the performance of modern codes with their high complexity of decoding algorithms represents a notable gap to the normal approximation at short block systems. Due to this reason, the studies in [6,13], and [31–33] have reviewed and compared channel coding schemes and their construction methods to cope with short packet transmission systems.

The surveys provided in [9,34] about channel codes in MTC are relevant to scenarios which require moderate performance with PER of $10^{-4}$ and few ms latency. While turbo codes, LDPC, and polar codes offer similar performance at long packet transmission, they have a considerably different performance at short packet lengths which should be further improved to support MTC [15].

In latency comparison, CCs outperform other codes as the CCs start encoding/decoding immediately upon receiving bits. In contrast, polar codes and LDPC codes need to receive all block bits to start encoding/decoding. The remarkable reliability performance of polar codes and LDPC codes as compared to CCs and turbo codes make them widely researched. However, many studies have addressed the performance superiority of polar codes as compared to LDPC codes in short block lengths without any sign of error floor [30]. For example, polar codes with Successive Cancellation List (SCL) decoding algorithm make a $10^{-4}$ reliability available with just a 0.5 dB gap to the normal approximation with noticeable complexity [35]. In general, polar codes outperform turbo codes, CCs, and LDPC codes in terms of time complexity, reliability, and tradeoffs [36,37]. However, 5G has already chosen polar codes and LDPC for Mobile Broadband (eMBB) control and data channels respectively. Despite that, the existing candidate channel codes for mcMTC still indicate a significant gap to the benchmark of the normal approximation performed by [38], hence require further improvement.

This section shows that mcMTC can work through the IEEE 802.11 standard with the improvement needed in the PHY layer level which basically requires ultra-reliable channel coding scheme to work on short block length without error floors. Hence, our motivation for this research lies between the considerable gap of the industrial wireless performance solutions that recently exist in the marketplace and literature, and the stringent requirements of various mcMTC deployment scenarios.

## 3 Low-Latency PHY Layer

In this section, we will discuss some of the latency enablers on the PHY layer level such as short packet transmission, the bandwidth, and using a simple and low complex modulation scheme. We will also review various IEEE 802.11 versions in providing low latency communication by means of packet transmission time in order to evaluate the performance of proposed PHY layer for mcMTC.

### 3.1 Short Packet Transmission System

Packet length can be clearly defined and formulated based on the definition of packet transmission as it is the process of transmitting information bits over the wireless channel after mapping them into a continuous-time signal. Packet length is thus defined as the number of channel uses that are required to transfer information bits. The packet length $n$ is given by $n \approx BT$, here $T$ denotes the approximate duration of continuous-time signal, and $B$ indicates approximate channel bandwidth.

Practically, current wireless networks are designed to work on long packet transmission. Logically, when the size of packet is long, the distortions and thermal noise introduced by the propagation channel are averaged out due to the large numbers' law. In the case of MTC with a short packet, such averaging fails. The terminology assigned to the different packet sizes is given by [39] and has been represented in Tab. 1.

**Table 1:** Packet length terminology

| Packet Length (L) in bits | Description |
| --- | --- |
| L < 100 | Very short |
| $100 \geq L < 600$ | short |
| $600 \geq L < 1024$ | Medium |
| $L \geq 1024$ | long |

The Fig. 1 illustrates the steps of packet creation in the communications system, where the information bits $k_i$ and control information $k_o$ form the payload of the packet. The payload bits are encoded and extended into $n_e$ complex numbers in order to improve the reliability of packet transmission. Lastly, extra symbols $n_o$ are appended to allow the receiver to compensate any distorted signals transmitted via a wireless channel.

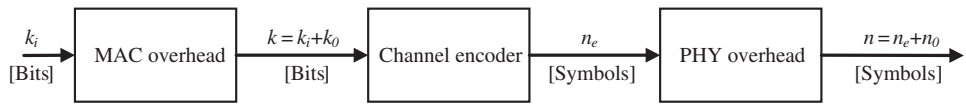

**Figure 1:** Block diagram of packet creation process

In the short packet communication systems, higher spatial diversity (which is measured by the transmission rate) is more preferable than spatial multiplexing. The transmission rate $R$ is measured by the number of transmitted payload bits per second per unit bandwidth. It is given by $k/n$ where $K$ denotes the information bits and $n$ represents the number of transmitted symbols. However, mcMTC desires higher transmission rate for latency reduction without violating the desired reliability (see subsection 3.2). The normal approximation given in [14] suggests that in sustaining the required reliability for a particular size of packet $n$, one encounters a rate $1/\sqrt{n}$ penalty comparison. However, with its short packet communication toolbox "spectre", we can approximate the packet length under PER and SNR constraints. For example, to operate at 60% of channel capacity with PER of $10^{-5}$ and SNR of 3 dB, it is sufficient to use codes in which packet size is greater than 300 bits (see Fig. 2). It also shows that with PER of $10^{-5}$ and SNR of 3 dB, the maximum channel capacity that can be achieved is 80%.

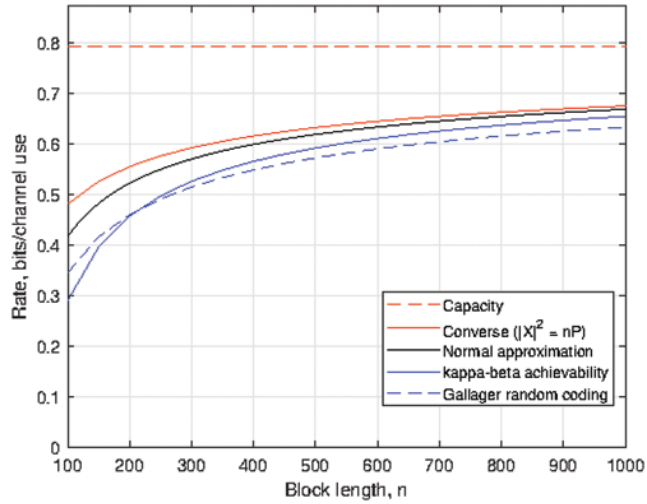

**Figure 2:** Bounds for AWGN, SNR = 3 dB, PER = $10^{-5}$

### 3.2 Reliability or Low-Latency Phase Modulation

Modulation and coding are traditional techniques used to increase transmission rate and decrease PER. In general, for sufficiently strong wireless links, higher modulation order together with light coding scheme can substantially boost transmission rate and decrease transmission delay [40]. However, when the channel becomes deficient, transmission rate should be reduced through the adoption of lower orders of modulation or more robust coding scheme to maintain the target PER. Indeed, high–order modulations and code rates allow very high data rates but suffer from poor reliability and may be unnecessary if the number of bits per packet is low. In this case, Binary Phase Shift Keying (BPSK) is a good modulation scheme to be used.

BPSK refers to a scheme of modulation in two phases where the binary messages of 0's and 1's are represented by two separate phase states in the carrier signal, separated by 180°. BPSK is the simplest and strongest form of Phase Shift Keying (PSK) because it needs the highest level of noise to produce incorrect decision in the demodulator [6]. This technique of modulation permits partial reuse of existing designs such as the IEEE 802.11 standard, easy compliance to the regulations of the spectrum mask, and easier channel equalization. Due to the low BPSK modulation which only modulates at one bit per sample, BPSK superior other OFDM modulations in terms of covered range [30] and reliability (see subsection 6.2).

### 3.3 Latency Aware IEEE 802.11 Standard

The PHY layer latency $T$ in terms of packet transmission time is basically depends on the payload size $L$ in bits, transmission rate $R$ in bit per second (bps), bandwidth $B$ in Hz, and data rate $S$ in bits per channel use (bpcu). Therefore, the latency $L$ in seconds is mathematically represented as:

$$T = \frac{L}{2BS} = \frac{L}{R} \tag{1}$$

From Eq. (1), the latency can be regulated by the selection of the packet size and bandwidth. The higher the available transmitting bandwidth and the lower of packet size, the lower the latency achieved. In such transmission regime, the spectral efficiency reduces because of the resulting low data rate. However, increasing bandwidth is a straightforward approach of providing diversity to the mcMTC transmissions [41] (see Fig. 3).

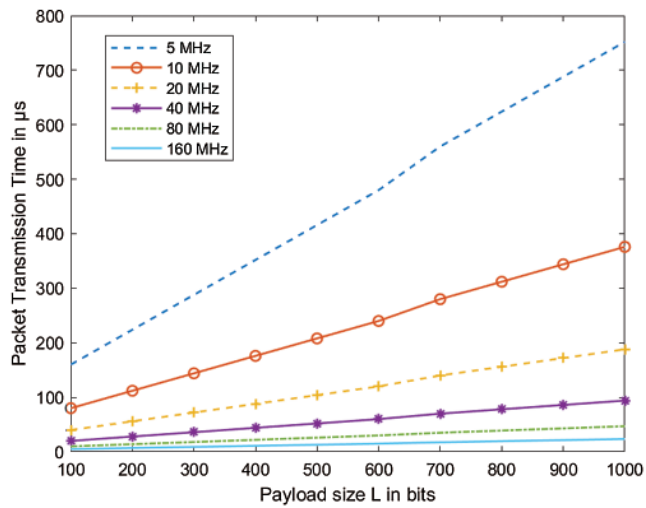

**Figure 3:** Packet transmission time versus. payload size and bandwidth for 802.11 OFDM with BPSK modulation and code rate at 1/2 over AWGN

Tab. 2 depicts the packet transmission time in various IEEE 802.11 standards using BPSK modulation for transferring 100 bits and assuming absence of channel coding and only one spatial stream is used. Although the most recent IEEE 802.11 standards assign much larger bandwidth near 160 MHz, their packet transmission time is higher than the legacy IEEE 802.11a/g standard. Practically, the duration of preamble exceeds the packet transmission time in the most critical scenarios (including synchronization margins, acknowledgement, data transmission, etc.).

In general, all new standards are not designed to realize short packet transmission such that the PHY layer overheads are suboptimal for short packet transmission [8]. Thus, they fail to lower the duration of the packet below the suggested value (20 μs) to handle the advanced mcMTC control applications [42]. Therefore, the study in [8] is based on the IEEE 802.11a PHY layer infrastructure and it is also selected as a base work in this paper for mcMTC. However, the PHY layer design of IEEE 802.11a requires optimization for the inefficiencies affecting short packet transmission (which will be the task of our next research work). However, the current IEEE 802.11a standard necessitates SNR of 8 dB to reach PER at $10^{-4}$.

**Table 2:** Transmission time for various IEEE 802.11 standards

| Standard | $B$ (MHz) | $N_{FFT}$ | $N_{ds}$ | $N_{cp}$ | $N_{ps}$ | $T_{pkt}$ ($\mu$s) |
|---|---|---|---|---|---|---|
| IEEE 802.11a/g | 20 | 64 | 48 | 16 | 5 | 32 |
| IEEE 802.11n | 20 | 64 | 52 | 16 | 7 | 40 |
| | 40 | 128 | 108 | 32 | 7 | 36 |
| IEEE 802.11ac | 80 | 256 | 234 | 64 | 10 | 44 |
| | 160 | 512 | 468 | 128 | 10 | 44 |

### 3.4 Compatibility of IEEE 802.11 Standard for MTC

Many efforts have been dedicated in adopting wireless networks, including cellular networks for MTC as discussed in Section 2. However, wireless local area networks (WLANs) are considered as interesting alternatives to cellular networks in supporting mcMTC as they:

1. Provide high data rate with high adaptability and scalability [27].
2. Offer more cost-effective solutions and provide longer transmission range.
3. Possess the features of group-centric communications, small data transmission, and low mobility [43].
4. Do not depend on cellular service providers.
5. Can support mcMTC with lesser efforts in adopting the PHY layer as some of them already have fewer OFDM symbols and shorter preamble duration.
6. Can cope with the applications that require high refresh rates or periodic mcMTC [27], while cellular networks cannot.

## 4 Ultra-reliability Channel Coding for mcMTC

Reliability refers to the probability of transmitting information bits from one network node to another within a given latency bound. Reliability, as a basic problem, has been properly studied in the theory of information [44]. Therefore, it has become a research hotspot from the perspective of networking to measure the performance of used PHY layer techniques. MTC over wireless networks faces challenges in reliability due to wireless tampering and sniffing, intermediate routing, and long-distance transmissions. Increasing reliability of every transmitted packet leads to minimized retransmissions and reduced latency. The key enablers for improving reliability in wireless communications have been listed in [45].

Two approaches can be applied to improve PHY layer packet transmission reliability. The first approach is the automatic repeat request technique, wherein several check bits are added to a packet to check whether the packet is received without errors. Although this technique can meet stringent PER requirements, the retransmitting mechanism causes a long delay. Thus, it is impractical for mcMTC. Another approach is FEC which is widely employed in modern wireless communication standards and has been considered in this paper.

### 4.1 Channel Codes for Short Block Length MTC

Channel coding is an essential part of PHY layer in communication systems. A channel code denotes a map that exists between the information payload transmitted as signals over the air and the information payload received by the receiver which is recovered from a noisy and distorted version of transmitted signal. Fig. 4 presents a diagram of a simplified communication system. Polar codes and LDPC codes are block-wise wherein the entire sequence must be available at the input before starting their work, inherently causing a delay. Therefore, the messages are divided into blocks for encoding and decoding to be done separately. Consequently, the error correction performance degrades with decreased block lengths and the adoption of an efficient channel code becomes critical. Shannon's channel coding theorem performs well over infinite block lengths with a vanishing error probability. Unfortunately, this is mostly impractical in short block lengths. Modern channel codes such as CCs, polar codes and LDPC codes show promising performance for short block lengths and various rates of code. Thus, they are assessed for mcMTC [46].

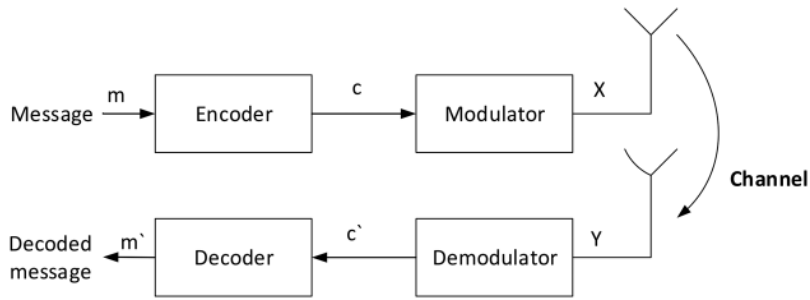

**Figure 4:** Channel coding/decoding in communication system

#### 4.1.1 Convolution Codes (CCs)

CCs are error-correcting coding technique first introduced by Elias in 1955. CCs are characterized by three parameters: the input length $k$, the memory elements number $m$, and the output length $n$. These codes can be processed while receiving bits as they contain memory, distinguishing them from block codes. Two popular Convolutional decoders are frequently used: the soft input Viterbi algorithm (VA) and tail-biting CC (TBCC). While the former requires extra zeros at the short block end in which its length is extended by a significant percentage, the complexity of the latter is observed to be $S$ times larger than VA [47], where $S$ represents the states number in the trellis.

#### 4.1.2 LDPC Codes

LDPC codes were first discovered by Gallager in 1962 and were later rediscovered in 1993. LDPC codes are type of error-correcting code in linear blocks with low-density sparse parity-check matrix and decoding algorithms that are practical and simple. Due to the iterative nature of LDPC decoding algorithms, there performance depend on the iterations which are decided in advance according to the system requirement. There has been growing interests in LDPC codes both in the industry and in academia as their performance is close to the Shannon limit [48] and they are applicable to all channels and code lengths. Therefore, LDPC codes are adopted for numerous communication systems like 802.11n, 802.16e, 5G, and mcMTC [15]. The iterative belief propagation (BP) decoding algorithm is commonly used for LDPC decodes.

### 4.1.3 Polar Codes

In the theory of coding, polar codes have attracted vast research attention since being introduced by Arıkan in 2008 [49], especially in code construction and decoding algorithms [50]. However, polar codes are the initial codes for practical use which can provably obtain the capacity of the channel at an infinite length. Furthermore, they are suitable for a variety of diverse condition scenarios because of their encoder's capability to function with the various kinds of decoders. Channel polarization is a central technique used in constructing polar codes, which consists of finding the most unreliable set of bit positions, commonly known as frozen set. The rest of the set are utilized for transmitting information bits. Since polar codes perform better than all current channel codes, particularly for short code length [51], they are considered very promising in mcMTC.

Decoding in polar codes can be performed in several ways, typically using SCL and SC decoding [51]. SC decoding algorithm has a complexity of $O(N \log N)$, and PER exponentially decays in the square root of large code word lengths. However, SC decoding algorithms gain latency challenge due to their recursive nature. Therefore, SC is not suitable for short-to-medium block lengths compared to SCL decoding algorithm which enhances polar codes performance at all block lengths [52]. The increase in list size of SCL decoder enhances the reliability performance of polar codes which however increases the implementation complexity [15]. The flexibility of SCL allows the tradeoff exploration between latency decoding performance, reliability, and through-put [51,53]. The Cyclic Redundancy Check (CRC) is a noticeable enhancement to SCL, in which a CRC error detection code encodes the message, resulting in polar codes. The checksum of CRC is utilized at the end of the decoder to select the correct path for list decoding. The comparison of CCs, LDPC codes, and polar codes is summarized in Tab. 3.

**Table 3:** Comparison between CCs, LDPC, and polar codes channel coding schemes

| Parameter | CCs | LDPC | Polar codes |
|---|---|---|---|
| Computational complexity | Low for low coding rate | High for low coding rate | Low for most coding rate |
| Decoding algorithm | TB, VA | BP, Log-BP | SC, SCL, CRC-SCL |
| Energy efficiency | High for low coding rate | Low at lower coding rate | High for all coding rate |
| Hardware efficiency | Low | Higher | Lower |
| Throughput | Low | High for higher coding rate | High for pipeline structure |
| Has error floor issue | No | Yes | No |

### 4.2 Compatibility of Polar Codes for mcMTC

In the selection of channel coding in mcMTC, polar codes are considered the most acceptable contender for many reasons listed below:

1. The reliability performance of polar codes outperforms other codes at low code rates for short block transmission system with no signs of error floor [30].

2. In comparison to other code candidates for the same error rate, polar codes require lower SNR. Thus, polar codes improve spectral efficiency with higher coding gain.

3. The utilization of low complexity SC-based algorithms causes polar decoders to consume lesser power, which is almost 20 times lesser than that of turbo codes with equal complexity [46].

4. Polar codes satisfy the demands for low latency in mcMTC [54].

Due to the advantages mentioned above, in October 2016, polar codes became the dominant channel coding scheme for Chinese Huawei 5G communication. Huawei announced that polar codes could achieve the speed of 27 Gbps. Since then, polar codes have been selected for controlling channels of short blocks in 5G eMBB, causing polar codes to be widely researched for supporting mcMTC.

## 5 Ultra-reliable PHY Layer Design for mcMTC

In this section, we will discuss the proposed PHY layer design over wireless network which is based on IEEE 802.11a standard with the following amendments:

1. We have proposed for the first time to implement polar codes in legacy IEEE 802.11 standard aiming at providing ultra-reliable mcMTC for real-time applications.

2. We considered using BPSK modulation due to its low complexity, and high reliability which surpass the other modulations.

3. We considered short packet length structure in our implementation for mcMTC.

4. The new proposed PHY layer works on bandwidths at 20 MHz and 80 MHz to provide latency over transmission link.

Implementation of polar codes in OFDM based PHY layer requires restructuring OFDM transceiver from a bit-coding to a block-coding architecture. In transmitter with CCs, the message bits are divided into OFDM symbols, each of which is encoded separately (see Fig. 5a). Similarly, the CCs in the receiver start decoding once they are completely received and continue decoding until all symbols are received. In contrast, polar codes in the transmitter apply encoding to the entire message bits and then divide them into OFDM symbols (see Algo. 1). Similarly, polar codes in the receiver will wait until all OFDM symbols are received to start their decoding (see Algo. 2). Even though, polar codes waiting cause some delay, the system encoding/decoding calls are only two time in case of using polar codes and $2N_{sym}$ (number of the OFDM symbols) in case of using CCs. For example, transmitting 128 message bits in IEEE 802.11a requires 6 symbols with BPSK modulation ($mod$), code rate ($R_c$) at 0.5, and 48 data subcarrier ($ds$). Hence CCs will be called 12 times and polar codes will be called two times (see Fig. 5).

---

**Algorithm 1:** The OFDM transmitter with polar codes

---

**Input:** $<m, R_c, ds, N_{sym}, mod>$
**Output:** $Tx_{wf}$
$n \leftarrow$ pc_encode($m$)
$n' \leftarrow$ add_padding_bits($n$)
$nm \leftarrow$ modulate($x, mod$)
$s \leftarrow 0$; data $\leftarrow$ []
**While** $s < N_{sym}$ **Do**
    $x \leftarrow nm \ (s \times ds + 1 \rightarrow (s+1) \times ds)$
    $xf \leftarrow$ Ifft($x$)
    $xcp \leftarrow$ Add_cp($xf$)
    $data(s+1) \leftarrow xcp$
**End**
$Tx_{wf} \leftarrow data$

---

**Algorithm 2:** The OFDM receiver with polar codes

---

**Input:** $<Rx_{wf}, m, R_c, ds, N_{sym}, mod>$
**Output:** $m'$ // decoded message bits
$dy \leftarrow$ demodulate($Rx_{wf}$)
$s \leftarrow 0$; data $\leftarrow$ []
**While** $s < N_{sym}$ **Do**
    $y \leftarrow dsy \ (s \times ds + 1 \rightarrow (s+1) \times ds)$
    $yf \leftarrow$ fft($y$)
    $ycp \leftarrow$ remove_cp($yf$)
    $data(s+1) \leftarrow ycp$
**End**
$md \leftarrow$ Pc_decode($data$)
$m' \leftarrow$ remove_padding_bit s()

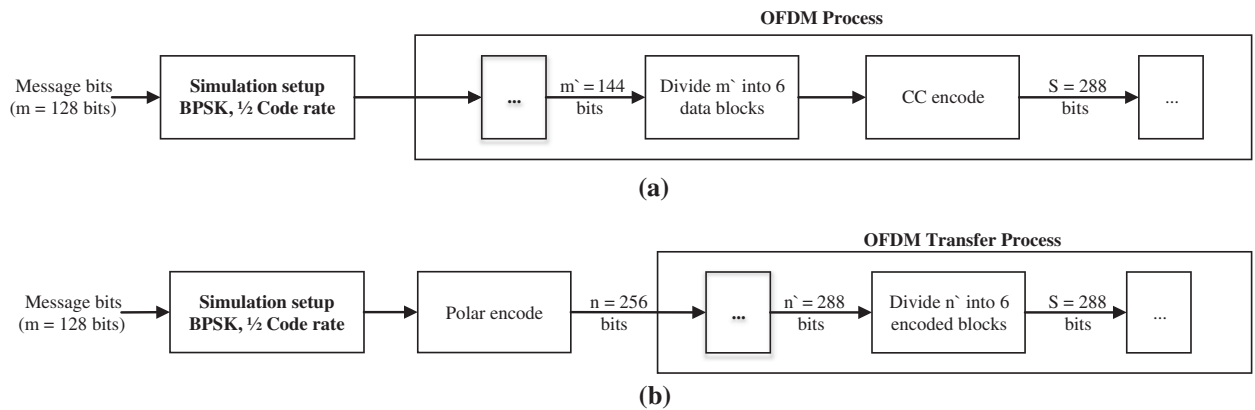

**Figure 5:** PHY layer simplified diagram of encoding 128 message bits by using CCs versus polar codes considering BPSK modulation and code rate at 1/2. (a) CC encoding process (b) Polar encoding process

## 6 Performance Evaluation of Proposed mcMTC PHY Layer

In this section, we will elaborate the reliability performance of the proposed PHY layer. For this purpose, we have evaluated the performance of different channel codes by means of BER and PER on the condition of short packet lengths with BPSK modulation using Monte Carlo simulation at different levels of SNR for a variety of payload lengths. AWGN channel has been selected in this simulation as a normal channel, which can reflect the real signal transmission conditions in the best way.

### 6.1 Simulation Model

The system setup considers only the PHY layer design in a network protocol stack based on the IEEE 802.11a standard. The received message $k$ is encoded into $n$ bits. It then goes through OFDM process including BPSK modulation. After that, the signals are ready to be transmitted via AWGN channel with given SNR noise in EbN0. At the receiver, the reverse operation is applied, and the received message is decoded and tested for errors in terms of BER/PER. Finally, BER and PER are plotted against EbN0. Fig. 6 shows the simulation model followed in this paper and Tab. 4 summarizes the simulation parameters used in the simulation. VA, Log-BP, and CRC-SCL decoding algorithms have been adopted to CCs, LDPC codes, and polar codes, respectively.

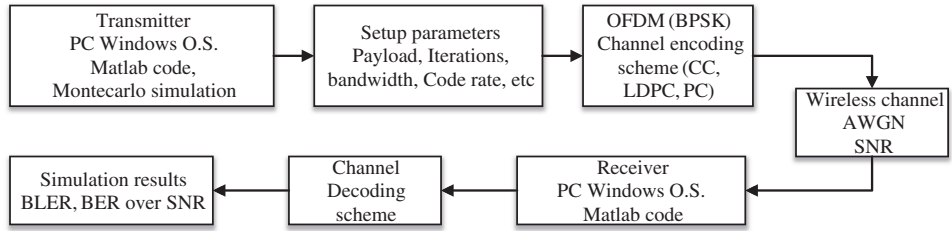

**Figure 6:** System model

**Table 4:** Simulation parameters and specifications

| Parameters | Specifications |
|---|---|
| Multiplexing technique, Modulation scheme | OFDM, BPSK |
| Channel model | AWGN |
| Channel code | CC, LDPC, Polar |
| Design SNR dB | 2 dB |
| Decoding algorithm | VB, BP, CRC-SCL |
| Polar code construction method | Bhattacharyya parameter |
| CRC, SCL | 16, 16 bits |
| BP iterations | 200 |
| Code length | 64, 128, 256, 512 bits |
| Code rate | 1/2 |
| Simulation iterations | $10^6$ |
| Minimum BER to stop iteration | 100 BER |

### 6.2 Performance Evaluation of OFDM Modulation Schemes

The simulation was performed based on IEEE 802.11a for transferring 256 message bits for different modulation schemes with no channel coding. The simulation results depicted in Fig. 7 show that the reliability performance of BPSK outperforms the others as it requires SNR to be 10 dB to reach zero error probability. However, QPSK, 16-QAM, and 64-QAM require 12 dB, 16 dB, and 28 dB respectively to reach the same BPSK results. Therefore, BPSK has been selected as the best modulation option for low-latency mcMTC in short packet length regime.

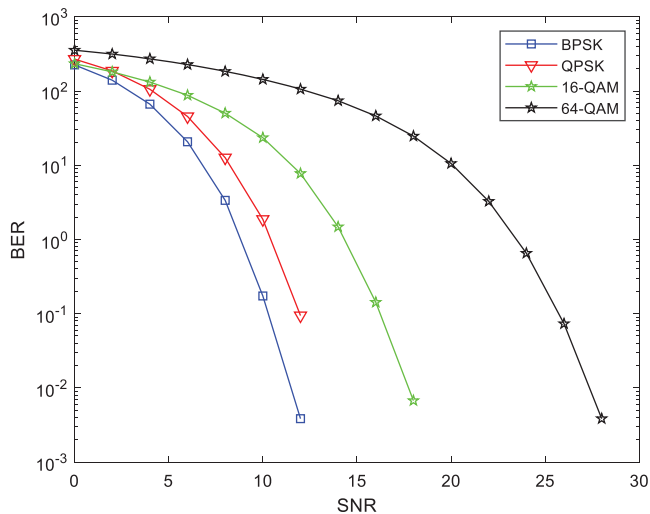

**Figure 7:** Short packet OFDM modulation reliability performance over AWGN channel

### 6.3 Performance Evaluation of Channel Coding Schemes

The simulations conducted to measure the reliability performance for different channel coding schemes over AWGN are reported here.

Fig. 8 shows that in block size at 512 bits, polar codes outperform others for SNR greater than 1 dB such that polar codes can reach $10^{-5}$ with SNR at 2.5 dB and $10^{-7}$ with SNR at 3 dB with 1 dB gap to the normal approximation. While LDPC codes performance is almost 1 dB far from polar codes the PER performance of LDPC codes outperform polar codes in SNR lower than 1 dB. The CCs performance is far away from both LDPC and polar codes.

In case of using block size at 256, Fig. 9 shows that both PER and BER of polar codes can reach zero error at almost SNR = 3 dB with 1 dB gap to the normal approximation. Comparatively, LDPC codes require SNR at 4.5 dB to get zero BER and more than 5 dB to reach zero PER. Again, CCs show the worst performance.

The results given in Figs. 10–11 show that LDPC codes outperform Polar codes for packet length at 128 bits and 64 bits with SNR ≤ 1.5 dB and SNR ≤ 2.5 dB respectively. However, polar codes compete LDPC codes for higher SNR values. The results also show that with packet lengths lower than 128 bits, the polar codes performance gap to the normal approximation is reduced to be 0.5 dB.

Monte Carlo simulation reveals interesting results. It is evident that polar codes using CRC-SCL decoding algorithm outperform others in all block lengths at a fixed code rate of 1/2 and have the minimum gap to the normal approximation. This is due to the polar code construction and decoding capability of the CRC-SCL decoding algorithm. The results also confirm the supremacy of polar codes at a high SNR regime. In order to satisfy the minimum reliability requirement at BER $10^{-5}$, polar codes need 2.3 dB, 2.7 dB, 3.5 dB, and 4 dB for block lengths at K = 512 bits, K = 256 bits, K = 128 bits, and K = 64 bits, respectively. Similarly, in order to reach the same reliability, LDPC codes need 3.5 dB, 4.1 dB, 5 dB, and greater than 5 dB for block lengths at K = 512 bits, K = 256 bits, K = 128 bits, and K = 64 bits, respectively. Based on the discussed results, we can conclude that adopting polar codes instead of CCs in IEEE 802.11 standard will reduce the gap to the normal approximation by more than 4 dB.

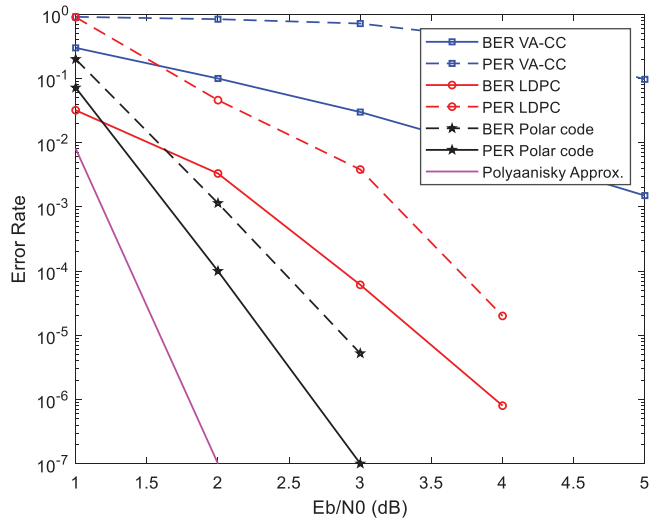

**Figure 8:** Error rate *vs.* EbN0 for K = 512 bits

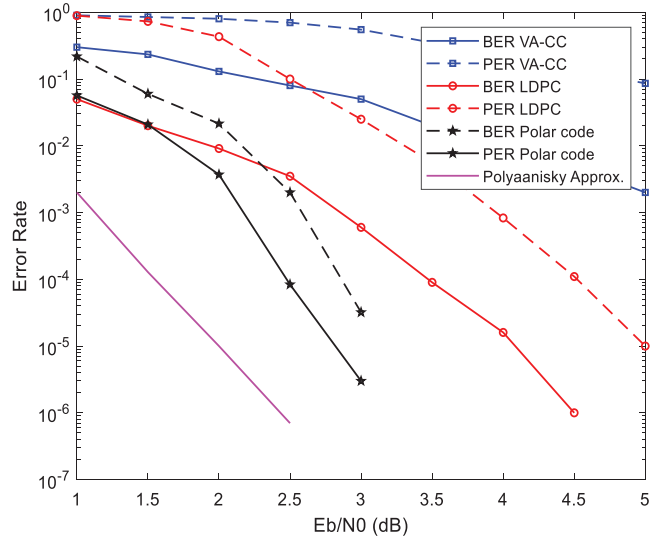

**Figure 9:** Error rate *vs.* EbN0 for K = 256 bits

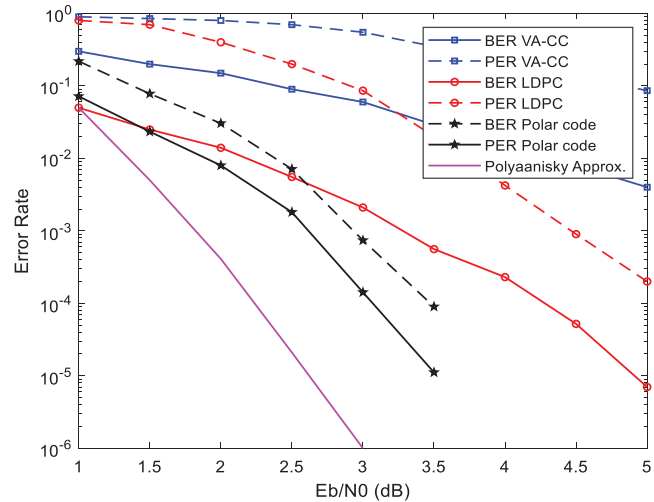

**Figure 10:** Error rate *vs.* EbN0 for K = 128 bits

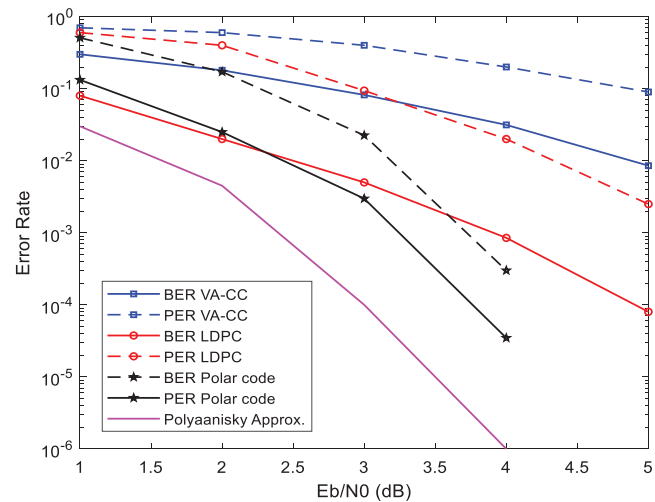

**Figure 11:** Error rate *vs.* EbN0 for K = 64 bits

### 6.4 *Errors Classification of Transmitted Packet*

During the simulation process, we noticed that it is beneficial to provide a classification to the bit errors based on the source of the error. The bit errors are classified into three groups: the packet detection error $P_{det}$, the signal error $S_{err}$, and the data decoding error $D_{err}$. Fig. 12 represents these errors for both IEEE 802.11a with CCs (CCs-IEEE 802.11a) and IEEE 802.11a with polar codes (PC-IEEE 802.11a). The figure shows that the data decoding errors form 75% to 90% of the total errors detected in CC-IEEE 802.11. This degradation in data decoding performance is due to its low CCs performance. In PC-IEEE 802.11a, the majority of the errors are signal detection errors which form 63% to 83% of the total errors detected. No errors were detected for polar codes with SNR 6 and 7. The improvement in decoding packets is return to the use of polar codes. Based on given results, adopting polar codes in IEEE 802.11a is prominent in realizing the latency and reliability requirements of mcMTC.

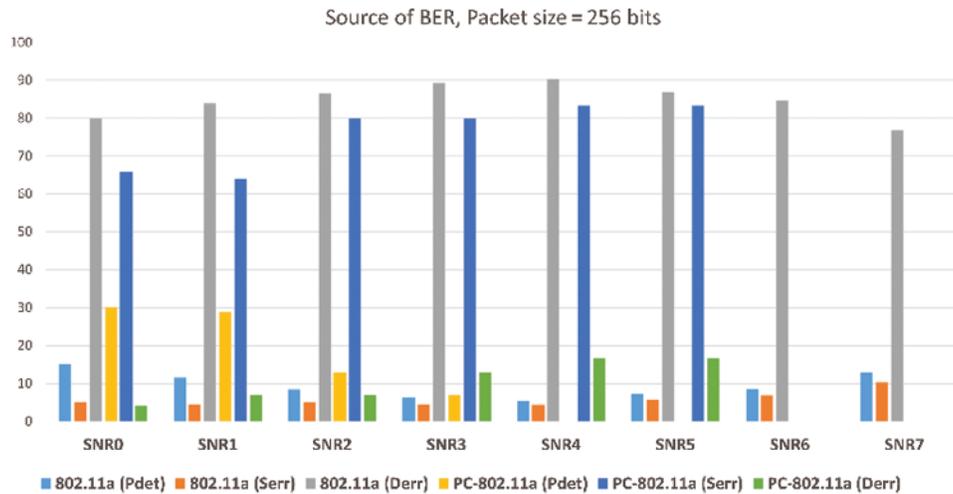

**Figure 12:** Source of BER errors of IEEE 802.11a and Polar codes IEEE 802.11a for K = 256 bits

## 7  Conclusion and Future Work

This research work implements polar codes as the FEC scheme instead of CCs FEC in the OFDM based PHY layer of wireless networks with the aim of providing ultra-high reliability and low latency through using of highly reliable BPSK modulation and short packet size. The proposed PHY layer design follows the IEEE 802.11a standard to ensure using fewer OFDM symbols. The proposed design is validated through simulation which is performed in MATLAB to investigate and compare polar codes, LDPC codes, and CCs performance in short packet transmissions. The simulation results show that the reliability performance in terms of BER and PER versus SNR of polar codes with CRC-CSL decoding algorithm outperform the other codes in short packet transmission with a fixed code rate of 1/2 as they can provide reliability of BER at $10^{-5}$ with almost 0.5 dB gap to the normal approximation for transferring 128 message bits with reasonable complexity. Nonetheless, reduction in complexity may lead to better results. In this area of study, future research is needed for further reduction of PHY layer overhead, scrutiny of additional FEC codes, and optimization of the complexity/implementation of polar decoding algorithms to further latency and reliability improvement.

**Funding Statement:** The authors received no specific funding for this study.

**Conflicts of Interest:** The authors declare that they have no conflicts of interest to report regarding the present study.

## References

[1]  3GPP, "Service requirements for the 5g system," 3GPP TS 22.261, ver. 17.0.1, 2019.

[2]  M. Saad, M. Bin Ahmad, M. Asif, K. Masood and M. A. A. Ghamdi, "Social distancing and isolation management using machine-to-machine technologies to prevent pandemics," *CMC-Computers, Materials & Continua,* vol. 67, pp. 3545–3562, 2021.

[3]  A. Hammoodi, L. Audah, M. A. Taher, M. A. Mohammed, M. S. Aljumaily *et al.*, "Novel universal windowing multicarrier waveform for 5g systems," *CMC-Computers, Materials & Continua,* vol. 67, pp. 1523–1536, 2021.

[4]  M. Luvisotto, Z. Pang and D. Dzung, "Ultra high performance wireless control for critical applications: Challenges and directions," *IEEE Transactions on Industrial Informatics,* vol. 13, pp. 1448–1459, 2016.

[5] J. Åkerberg, M. Gidlund, F. Reichenbach and M. Björkman, "Measurements on an industrial wireless hart network supporting profisafe: A case study," *IEEE ETFA2011,* pp. 1–8, 2011.

[6] Y. Deng, M. Zhan, M. Wang, C. Yang, X. Luo *et al.*, "Comparing decoding performance of ldpc codes and convolutional codes for short packet transmissions," in *IEEE 17th International Conference on Industrial Informatics (INDIN)*, vol. 1, pp. 1751–1755, 2019.

[7] Y. F. Huang and H. H. Chen, "Physical layer architectures for machine type communication networks a survey," *Wireless Communications and Mobile Computing,* vol. 16, pp. 3269–3294, 2016.

[8] X. Jiang, Z. Pang, M. Zhan, D. Dzung, M. Luvisotto *et al.*, "Packet detection by a single ofdm symbol in urllc for critical industrial control: A realistic study," *IEEE Journal on Selected Areas in Communications,* vol. 37, pp. 933–946, 2019.

[9] M. Zhan, Z. Pang, M. Xiao, M. Luvisotto and D. Dzung, "Wireless high-performance communications: Improving effectiveness and creating ultra-high reliability with channel coding," *IEEE Industrial Electronics Magazine,* vol. 12, pp. 32–37, 2018.

[10] I. PAS and 62601: "Industrial networks-wireless communication network and communication profiles-wia-pa, " *International Electrotechnical Commission (IEC) Std,* 2015.

[11] Y.-H. Wei, Q. Leng, S. Han, A. K. Mok, W. Zhang *et al.*, "Rt-wifi: Real-time high-speed communication protocol for wireless cyber-physical control applications," in *IEEE 34th Real-Time Systems Symposium, IEEE*, pp. 140–149, 2013.

[12] D. Orfanus, R. Indergaard, G. Prytz and T. Wien, "Ethercat-based platform for distributed control in high-performance industrial applications," in *IEEE 18th Conf. on Emerging Technologies & Factory Automation (ETFA)*, pp. 1–8, 2013.

[13] G. Liva, L. Gaudio, T. Ninacs and T. Jerkovits, "Code design for short blocks: A survey," ArXiv Preprint ArXiv, no. 1610.00873, 2016.

[14] G. Durisi, T. Koch and P. Popovski, "'Toward massive, ultrareliable, and low-latency wireless communication with short packets," in *Proc. of the IEEE*, vol. 104, pp. 1711–1726, 2016.

[15] M. Shirvanimoghaddam, M. S. Mohammadi, R. Abbas, A. Minja, C. Yue *et al.*, "Short block-length codes for ultra-reliable low latency communications," *IEEE Communications Magazine,* vol. 57, pp. 130–137, 2018.

[16] S. Tong, D. Lin, A. Kavcic, B. Bai and L. Ping, "On short forward error-correcting codes for wireless communication systems," in *IEEE 16th Int. Conf. on Computer Communications and Networks*, pp. 391–396, 2007.

[17] D.-S. Yoo, W. E. Stark, K.-P. Yar and S.-J. Oh, "Coding and modulation for short packet transmission,", *IEEE Transactions on Vehicular Technology,* vol. 59, pp. 2104–2109, 2010.

[18] Y. Polyanskiy, H. V. Poor and S. Verduá, "Channel coding rate in the finite block length regime," *IEEE Transactions on Information Theory,* vol. 56, pp. 2307–2359, 2010.

[19] J. Åkerberg, M. Gidlund and M. Björkman, "Future research challenges in wireless sensor and actuator networks targeting industrial automation," in *9th IEEE Int. Conf. on Industrial Informatics*, pp. 410–415, 2011.

[20] H. C. Foundation, "Hart field communication protocol specification," HFC_SPEC12, revision 7.0, 2007.

[21] ISA, "Wireless systems for industrial automation: Process control and related applications," ISA-100.11, 2009.

[22] G. Scheible, D. Dzung, J. Endresen and J. E. Frey, "Unplugged but connected design and implementation of a truly wireless real-time sensor/actuator interface," *IEEE Industrial Electronics Magazine,* vol. 1, pp. 25–34, 2007.

[23] W. Liang, M. Zheng, J. Zhang, H. Shi, H. Yu *et al.*, "Wia-fa and its applications to digital factory: a wireless network solution for factory automation," in *Proc. of the IEEE*, vol. 107, pp. 1053–1073, 2019.

[24] F. Tramarin, A. K. Mok and S. Han, "Real-time and reliable industrial control over wireless lans: algorithms, protocols, and future directions," in *Proceedings of the IEEE*, vol. 107, pp. 1027–1052, 2019.

[25] X. Jiang, Z. Pang, M. Luvisotto, R. Candell, D. Dzung *et al.*, "Delay optimization for industrial wireless control systems based on channel characterization," *IEEE Transactions on Industrial Informatics,* vol. 16, pp. 5855–5865, 2020.

[26] A. Aijaz, "High-performance industrial wireless: Achieving reliable and deterministic connectivity over ieee 802.11 wlans," *IEEE Open Journal of the Industrial Electronics Society,* vol. 1, pp. 28–37, 2020.

[27] O. Seijo, Z. Fernández, I. Val and J. A. López-Fernández, "Sharp: Towards the integration of time-sensitive communications in legacy lan/wlan," *IEEE Globecom Workshops (GC Wkshps)*, pp. 1–7, 2018.

[28] A. Salh, L. Audah, Q. Abdullah, N. Shah, S. Hamzah *et al.*, "Energy-efficient low-complexity algorithm in 5g massive mimo systems," *Computers, Materials & Continua,* vol. 67, pp. 3189–3214, 2021.

[29] Q. Abdullah, N. Abdullah, A. Salh, L. Audah, N. Farah *et al.*, "Pilot contamination elimination for channel estimation with complete knowledge of large-scale fading in downlink massive mimo systems," 2021.

[30] G. T. R. W. Meeting, "Remaining issues on URLLC data channel coding," 3G*PP Technical Report*, 2018.

[31] W. K. Abdulwahab and A. A. Kadhim, "Comparative study of channel coding schemes for 5g," in *IEEE Int. Conf. on Advanced Science and Engineering (ICOASE)*, pp. 239–243, 2018.

[32] B. Tahir, S. Schwarz and M. Rupp, "Ber comparison between convolutional, turbo, ldpc, and polar codes," in *IEEE 24th International Conference on Telecommunications (ICT)*, pp. 1–7, 2017.

[33] P. Pathak and R. Bhatia, "Channel coding for wireless communication systems: A review," SSRN 3565791, 2020.

[34] A. Frotzscher, U. Wetzker, M. Bauer, M. Rentschler, M. Beyer *et al.*, "Requirements and current solutions of wireless communication in industrial automation," in *IEEE Int. Conf. on Communications Workshops (ICC)*, pp. 67–72, 2014.

[35] M. Sartipi and F. Fekri, "Source and channel coding in wireless sensor networks using ldpc codes," in *IEEE Communications Society Conference on Sensor and Ad Hoc Communications and Networks*, pp. 309–316, 2004.

[36] O. Iscan, D. Lentner and W. Xu, "A comparison of channel coding schemes for 5g short message transmission," in *IEEE Globecom Workshops (GC Wkshps)*, pp. 1–6, 2016.

[37] Z. R. M. Hajiyat, A. Sali, M. Mokhtar and F. Hashim, "Channel coding scheme for 5g mobile communication system for short length message transmission," *Wireless Personal Communications,* pp. 377–400, 2019.

[38] T. Erseghe, "Coding in the finite-blocklength regime: Bounds based on laplace integrals and their asymptotic approximations," *IEEE Transactions on Information Theory,* vol. 62, pp. 6854–6883, 2016.

[39] I. Corporation, "Channel coding scheme for URLLC, mMTC and control channels," 3GPP-TSG R1–167703, 2016.

[40] A. Marinšek, D. Delabie, L. De Strycker and L. Van der Perre, "Physical layer latency management mechanisms: A study for millimeterwave wi-fi," *Electronics,* vol. 10, no. 13, pp. 1599, 2021.

[41] P. Popovski, C. Stefanoviac, J. J. Nielsen, E. De Carvalho, M. Angjelichinoski *et al.*, "Wireless access in ultra-reliable low-latency communication (urllc)," *IEEE Transactions on Communications,* vol. 67, pp. 5783–5801, 2019.

[42] M. Luvisotto, Z. Pang and D. Dzung, "High-performance wireless networks for industrial control applications: new targets and feasibility," in *Proc. of the IEEE*, vol. 107, no. 6, pp. 1074–1093, 2019.

[43] 3GPP-TSG, "Service requirements for machine-type communications (MTC)," 3GPP-TSG TS22.368, 2012.

[44] C. E. Shannon, "A mathematical theory of communication," *The Bell System Technical Journal,* vol. 27, pp. 379–423, 1948.

[45] N. A. Mohammed, A. M. Mansoor and R. B. Ahmad, "Mission-critical machine-type communication: An overview and perspectives towards 5g," *IEEE Access,* vol. 7, pp. 127198–127216, 2019.

[46] A. Sharma and M. Salim, "Polar code appropriateness for ultra-reliable and low-latency use cases of 5g systems," *International Journal of Networked and Distributed Computing,* vol. 7, pp. 93–99, 2019.

[47] R. Wang, H. Xu, Y. Wei and D. N. Doan, "List viterbi decoding of tail biting convolutional codes," US Patent 8,543,895, 2013.

[48] M. Eroz, F.-W. Sun and L. N. Lee, "An innovative low-density parity-check code design with near-shannon-limit performance and simple implementation," *IEEE Transactions on Communications,* pp. 13–17, 2006.

[49] E. Arikan, "Channel polarization: A method for constructing capacity-achieving codes for symmetric binary-input memoryless channels," *IEEE Transactions on Information Theory,* vol. 55, pp. 3051–3073, 2009.

[50] C. Leroux, I. Tal, A. Vardy and W. J. Gross, "Hardware architectures for successive cancellation decoding of polar codes," in *IEEE Int. Conf. on Acoustics, Speech and Signal Processing (ICASSP),* pp. 1665–1668, 2011.

[51] I. Tal and A. Vardy, "List decoding of polar codes," *IEEE Transactions on Information Theory,* vol. 61, pp. 2213–2226, 2015.

[52] M. Maksimoviác and M. Forcan, "Application of 5g channel coding techniques in smart grid: ldpc vs. polar coding for command messaging," in *7th Int. Conf. on Electronics, Telecommunications, Computing, Automatics and Nuclear Engineering-IcETRAN,* pp. 746–751, 2020.

[53] M. Léonardon, A. Cassagne, C. Leroux, C. Jégo, L. P. Hamelin *et al.,* "Fast and flexible software polar list decoders," *Journal of Signal Processing Systems,* vol. 91, pp. 937–952, 2019.

[54] 3GPP TSG RAN WG1, "Channel coding schemes for URLLC scenario," In meeting no. 87, Huawei, HiSilicon, 2016.